\documentclass[pre,twocolumn, amsmath, amssymb, superscriptaddress]{revtex4}

\usepackage{graphicx}
\usepackage{dcolumn}
\usepackage{bm}
\usepackage{braket}
\usepackage[table]{xcolor}

\usepackage{amsmath}

\newcommand{\beq}{\begin{equation}}
\newcommand{\eeq}{\end{equation}}

\begin{document}

\title{In-Silico evidence for two receptors based strategy of SARS-CoV-2}



\author{Edoardo Milanetti\footnote{The authors contributed equally to the present work.}\footnote{Corresponding author: edoardo.milanetti@uniroma1.it}}
\affiliation{Department of Physics, Sapienza University, Piazzale Aldo Moro 5, 00185, Rome, Italy}
\affiliation{Center for Life Nanoscience, Istituto Italiano di Tecnologia, Viale Regina Elena 291,  00161, Rome, Italy}

\author{Mattia Miotto\footnotemark[1]}
\affiliation{Department of Physics, Sapienza University, Piazzale Aldo Moro 5, 00185, Rome, Italy}
\affiliation{Center for Life Nanoscience, Istituto Italiano di Tecnologia, Viale Regina Elena 291,  00161, Rome, Italy}

\author{Lorenzo Di Rienzo\footnotemark[1]}
\affiliation{Center for Life Nanoscience, Istituto Italiano di Tecnologia, Viale Regina Elena 291,  00161, Rome, Italy}

\author{Michele Monti}
\affiliation{Centre for Genomic Regulation (CRG), The Barcelona Institute for Science and Technology, Dr. Aiguader 88, 08003 Barcelona}
\affiliation{RNA System Biology Lab, Department of Neuroscience and Brain Technologies, Istituto Italiano di Tecnologia, Via Morego 30, 16163, Genoa, Italy}

\author{Giorgio Gosti}
\affiliation{Center for Life Nanoscience, Istituto Italiano di Tecnologia, Viale Regina Elena 291,  00161, Rome, Italy}

\author{Giancarlo Ruocco}
\affiliation{Department of Physics, Sapienza University, Piazzale Aldo Moro 5, 00185, Rome, Italy}
\affiliation{Center for Life Nanoscience, Istituto Italiano di Tecnologia, Viale Regina Elena 291,  00161, Rome, Italy}

\begin{abstract}
We propose a novel numerical method able to determine efficiently and effectively the relationship of complementarity between portions of protein surfaces. This innovative and general procedure, based on the representation of the molecular iso-electron density surface in terms of 2D Zernike polynomials, allows the rapid and quantitative assessment of the geometrical shape complementarity between interacting proteins, that was unfeasible with previous methods. 
We first tested the method with a large dataset of known protein complexes obtaining an overall area under the ROC curve of 0.76 in the blind recognition of binding sites and then applied it to investigate the features of the interaction between the Spike protein of SARS-CoV-2 and human cellular receptors. Our results indicate that SARS-CoV-2 uses a dual strategy: its spike protein could also interact with sialic acid receptors of the cells in the upper airways, in addition to the known interaction with Angiotensin-converting enzyme 2.
\end{abstract}

\maketitle

\section{Introduction}

At the time of writing, the new coronavirus, also known as SARS-CoV-2, which causes Severe Acute Respiratory Syndrome
\cite{huang2020clinical,zhu2020novel}, 
has caused approximately the death of approximately 17000  and the infection of 390000 people.
The COVID-19 outbreak represents a serious threat to public health~\cite{walls2020structure}, and the World Health Organization officially declared it a pandemic.

To date 7 coronavirus strains are known to infect humans and no approved therapies or vaccine against them are available~\cite{walls2020structure}.

In the past two decades, in addition to SARS-CoV-2, two other $\beta$-coronavirus have caused three of the most severe epidemics worldwide: SARS-CoV \cite{drosten2003identification, ksiazek2003novel} and MERS-CoV \cite{zaki2012isolation} which respectively cause the Severe Acute Respiratory Syndrome (SARS), and the Middle East Respiratory Syndrome (MERS).

The characteristics of the interactions between these viruses and the human cell receptors are being carefully studied to shed light on both diffusion speed and mortality rate differences between SARS-CoV-2 and the others, with special regard to SARS-CoV. 

Indeed, SARS-CoV spread across 26 countries in six continents and caused a total of 8,096 cases and 774 deaths (9.6\%) \cite{xu2020systematic}, with an incubation period of 1 to 4 days \cite{lessler2009incubation}. On the other side, it has been demonstrated that the latency of SARS-CoV-2 varies from 3-7 days on average, up to 14 days \cite{zhu2020novel}.
Thus, the average latency of SARS-CoV-2 is slightly longer than that of SARS-CoV \cite{xu2020systematic}.
Moreover, it is estimated from epidemiological data that individuals infected with SARS-CoV-2 are contagious from the beginning of the incubation period and that between the incubation period and the end of the infection each infected individual transmits the infection to about 3.77  other people \cite{yang2020epidemiological}.

SARS-CoV-2, similarly to SARS-CoV and MERS-CoV, attacks the lower respiratory system, thus causing viral pneumonia. However, this infection can also affect the gastrointestinal system, heart, kidney, liver, and central nervous system \cite{prompetchara2020immune, su2016epidemiology, zhu2020novel}. 

To face the emergency of this epidemic it is essential to shed light on the interaction mechanisms between the virus and the human cell receptors.

It is well characterized how SARS-CoV infection is mediated by the high-affinity interactions between the receptor-binding domain (RBD) of the spike (S) glycoprotein and the human-host Angiotensin-converting enzyme 2 (ACE2) receptor \cite{li2005structure,li2008structural,li2005receptor}. The spike protein is located on the virus envelope and promotes the attachment to the host cell and the fusion between the virus and the cellular membrane. \cite{graham2010recombination, kuo2000retargeting}.

Recently, it has also been proven that several critical residues in SARS-CoV-2's RBD provide favorable interactions with human ACE2, consistent with SARS-CoV-2's capability to infect the cell \cite{du2009spike,hoffmann2020sars}. 
On the experimental side, it has also been confirmed by in vivo experiments that SARS-CoV-2's entry is mediated by lung cell Ace2 receptors \cite{zhou2020pneumonia}.
More importantly, the structure of the spike-Ace2 receptor complex has been recently determined by Crio-em \cite{yan2020structural}.
In conclusion, it is now understood that SARS-CoV-2 binds to the ACE2 receptor to infect the host cell using its spike protein's RBD, even if it had most likely evolved from SARS-CoV independently \cite{andersen2020proximal}.

Since the ACE2 molecule is known to be a human entry receptor, the understanding of the molecular mechanisms of interaction between the ACE2 receptor and the spike protein of the virus can be a key factor designing new drug compounds. With this aim, computational methods based on both sequence and structure studies of proteins represent a powerful tool~\cite{wu2020analysis}.

Indeed, the development of effective computational methods for predicting the binding sites of proteins can improve the understanding of many molecular mechanisms \cite{donald2011algorithms,kortemme2004computational,gainza2020deciphering}. Several methods to analyze protein interaction have used protein surface information \cite{shulman2004recognition, duhovny2002efficient,sharp1990electrostatic,daberdaku2019antibody}.

A very promising strategy to study molecular interaction is to determine, using deep learning methods, chemico-physical features of the molecular surface \cite{gainza2020deciphering}. This method allows to efficiently detect binding sites but it, unfortunately, has the drawbacks of any other deep learning approach: it requires the definition and training of several parameters and the creation of a sufficiently large training and test database. This method also requires the analysis of all possible orientations and relative positions of the binding sites.

In this work, we present a new and general method to efficiently describe the shape of molecular surfaces and apply it to study the interactions between spike proteins and the corresponding receptors involved in the SARS-CoV, MERS-CoV, and SARS-CoV-2 infection mechanisms.

The method proposed here, for the first time, describes regions of a molecular surface with the 2D Zernike formalism. Indeed, each local patch of a 3D protein molecule can be represented as a surface on a 2D square grid, while retaining both distance and angular information with respect to a specific reference system. Applying the Zernike expansion we compute - completely unsupervised - numerical descriptors that summarize the patch geometry and we use them to compute the structural similarity between any two patches. The Zernike descriptors are rotationally invariant, ensuring to the method an high efficiency and a low computational cost since it avoids the necessity of any preliminary structure orientation. Overall, these operations substantially reduce the search space dimensionality, thus allowing our method to be able to explore several more protein regions compared to previous methods.

We first apply the method to a large dataset of protein-protein complexes (Protein-Protein Dataset), where we test the ability of our method to detect interacting regions from non-interacting regions with the use of our definition of complementarity. Our method recognizes interactions in the large Protein-Protein Database with good precision. Furthermore, thanks to the low computational cost of building an invariant description, we can blindly sample the entire surfaces of 2 interacting proteins, and retrieve their binding sites.

Then, we use our formalism to study the interaction between the spike protein and its membrane receptors, comparing SARS-Cov-2 with both SARS-COV and MERS. We demonstrate that the actual region of binding between spike protein and ACE2 human - both in SARS-CoV and SARS-CoV-2 - have a higher complementarity with respect to other randomly sampled exposed receptor regions.

Furthermore, we also analyze in detail the structural properties of the MERS spike protein which, like several other proteins belonging to coronaviruses family, can interact with sialic acids \cite{tortorici2019structural}. Among other coronaviruses, the bovine coronavirus (BCoV), and the two human coronaviruses OC43 and HKU1 employ sialoglycan-based receptors with 9-O-acetylated sialic acid (9-O-Ac-Sia) as key component \cite{hulswit2019human}. 
These viruses bind to cell surface components containing N-acetyl-9-O-acetylneuraminic acid and this interaction is essential for the initiation of an infection \cite{schwegmann2006sialic}. In particular, we propose here a possible alternative mechanism of SARS-CoV-2 cellular infection, through spike protein interaction with sialic acid receptors of the upper airways, similarly to what has been shown for the MERS spike protein \cite{park2019structures}. 
We indeed highlight that a surface region in the N-terminal domain of SARS-CoV-2 spike is very similar to the MERS spike sialic acid-binding region, and present a compatible charge. These observations suggest that these two pockets potentially share an analogous function. This second infection mechanism for  SARS-CoV-2 could explain its high diffusion speed.

\section{Results}

\begin{figure*}[]
\centering
\includegraphics[width = \textwidth]{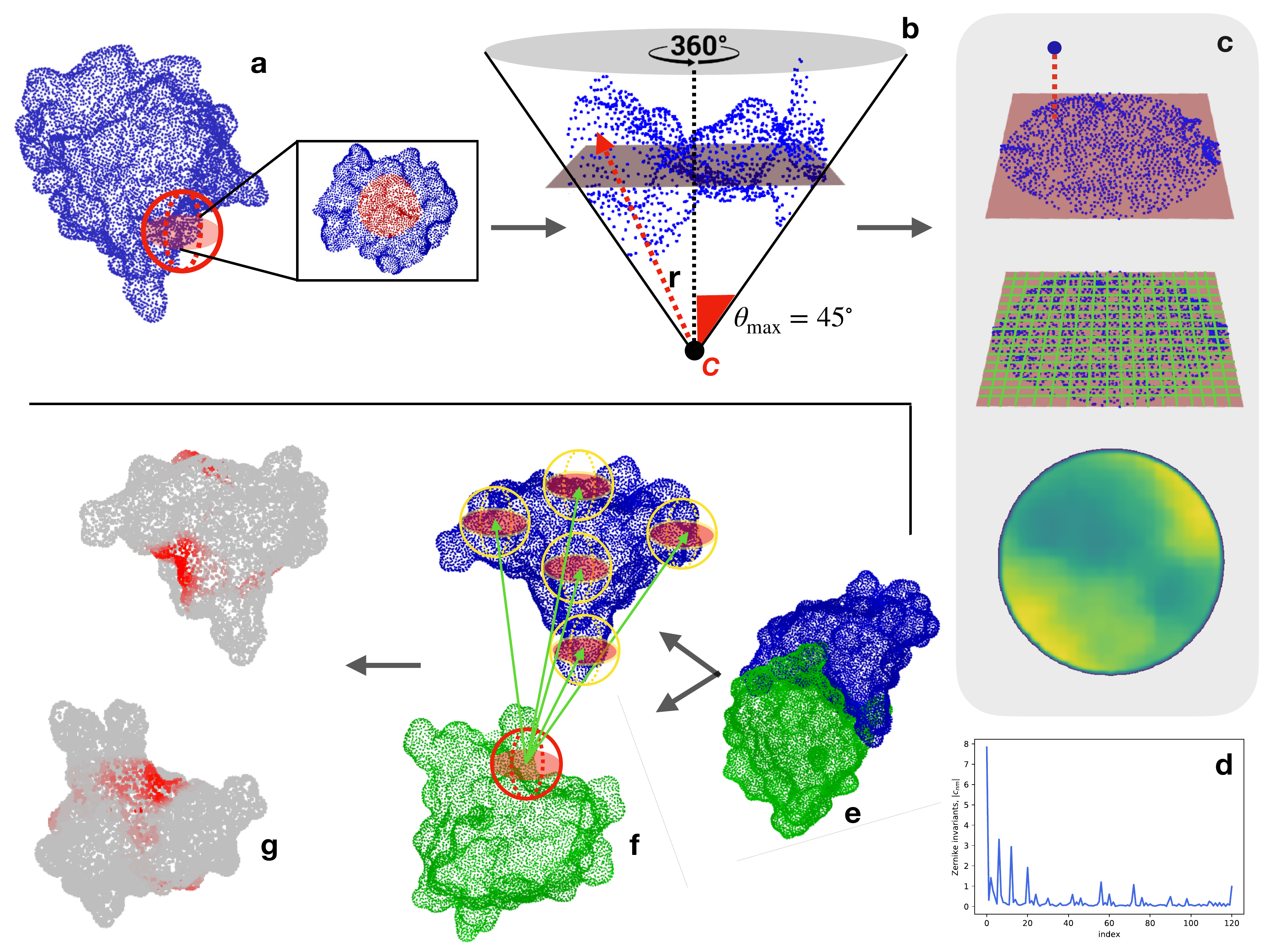}
\caption{\textbf{Computational protocol for the characterization of each surface region and the blind identification of the binding sites.}  \textbf{a)} Molecular solvent-accessible surface of a protein (in blue) and example of patch selection (red sphere). \textbf{b)} The selected patch points are fitted with a plane and reoriented in such a way that the z-axis passes through the centroid of the points and is orthogonal to the plane.  
A point C  along the z-axis is defined, such as that the largest angle between the perpendicular axis and the secant connecting C to a surface point is equal to $45^o$. Finally, to each point,  its distance, $r$ with point C is evaluated. 
\textbf{c)} Each point of the surface is projected on the fit plane, which is binned with a square grid. To each pixel,  the average of the r values of the points inside the pixel is associated. \textbf{d)} The resulting 2D projection of the patch can be represented by a set of 2D  Zernike invariant descriptors. \textbf{e-f)} Given a protein-protein complex (PDB code: 3B0F, in this example),  for each surface vertex we select a patch centered on it and compute its Zernike descriptors. To blindly identify the two binding sites, each sampled patch is compared with all the patches of the molecular partner and each vertex is associated with the minimum distance between its patch and all the patches of the molecular partner is associated with. \textbf{g)} A Smoothing process of the surface point values is applied to highlight the signal in the regions characterized mostly by low distance values (high shape complementarity).}
\label{fig:computational_protocol}
\end{figure*}

In the last decade, the 3D Zernike formalism has been widely applied for the characterization of molecular interactions \cite{venkatraman2009protein, kihara2011molecular, di2017superposition, daberdaku2019antibody}: in this work, we adopted a new representation, based on the 2D Zernike polynomials, which allows the quantitative characterization of protein surface regions. As shown in Fig.\ref{fig:computational_protocol}, our computational protocol associates to each molecular patch an ordered set of numbers (the expansion coefficients) that describes its shapes. 

Through this compact description, it is possible to both analyze the similarity between 2 different regions - suggesting, for example, a similar ligand for 2 binding regions - and to study the complementarity between 2 interacting surfaces.

To validate this method, we firstly collected a large structural dataset of protein-protein complexes and we characterize their binding sites through 2D Zernike. We test the method ability to recognize the higher complementarity observed between pairs of interacting surfaces compared to the lower complementarity found when the 2 surfaces are non-interacting. 

We thus analyze in detail the interactions of the SARS-CoV-2 spike protein with its membrane receptors, comparing SARS-CoV-2 properties with SARS-CoV and MERS-CoV.

\subsection*{Binding regions unsupervised recognition}

We selected a structural dataset, composed of about 4500 experimentally determined protein-protein complexes, from a recent paper that presented a state of the art patch recognition computational method \cite{gainza2020deciphering}.

For each complex, we have selected the interacting regions and we have characterized them with the 2D Zernike invariant descriptors. Therefore, each binding site is associated with a one-dimensional vector, allowing us to easily compare the shape of protein regions with the euclidean distance between their Zernike descriptors. Two regions are complementary when they are characterized by a low distance between their corresponding Zernike vectors\cite{venkatraman2009protein}.

To test the ability of the method to recognize two interacting regions, we have compared how much the distance between the Zernike descriptors (see Methods) of a pair of interacting binding sites is smaller than the distances between random patches. In particular, we define the random patch set as the set of 2000 patches randomly extracted from the 20 biggest protein of the dataset (100 patches for each protein). For each protein-protein complex, we define the real distance as the distance between the interacting surfaces, while the random ones are the values observed in the comparisons between the binding site of one protein and the random patch set.

Our {\it unsupervised} method has an excellent ability in recognizing the binding regions with respect to random patches with an Area Under the ROC Curve of 0.76. Note that the state of art {\it supervised} method described in \cite{gainza2020deciphering}, when only shape descriptors are considered, achieves an AUC of 0.68 with patches of comparable sizes, meaning that our approach clearly overcome such performances in shape complementarity recognition.

Furthermore, the much lower computational time needed for the calculation of the 2D Zernike descriptors allows an extensive sampling of the surfaces of a pair of proteins in complex. Centering on each surface point a molecular patch, we generate for each protein a very high number of Zernike descriptors. Comparing all the patches of the two proteins, we label each surface point with the \textit{binding propensity}, which is the maximum complementarity recorded between the Zernike descriptors of the patch and all the others belonging to the molecular partner surface. The real binding region is expected to be demarcated and mostly composed of elements with high complementarity. To make the binding region high complementarity more evident, we smooth the signal by attributing to each vertex of the surface the average value of the vertices closer than 6 $\AA$ to it (see Method).

As an example in Fig.\ref{fig:computational_protocol} we report the result of this method for a specific case (PDB code: 3B0F), where this procedure clearly identifies the binding regions of the two proteins. 
For a subset of 20 protein-protein complexes - the smallest in terms of the number of surface vertices -, we perform a blind search of the interacting regions. The results are very promising, since the average value of the AUC of the corresponding ROC is 0.65, with fourteen out of forty proteins having an AUC higher than 0.70.

\subsection*{Comparison between the complementarity of the SARS-CoV and SARS-CoV-2 spike protein with the human ACE2 receptor}
\begin{figure*}[]
\centering
\includegraphics[width = \textwidth]{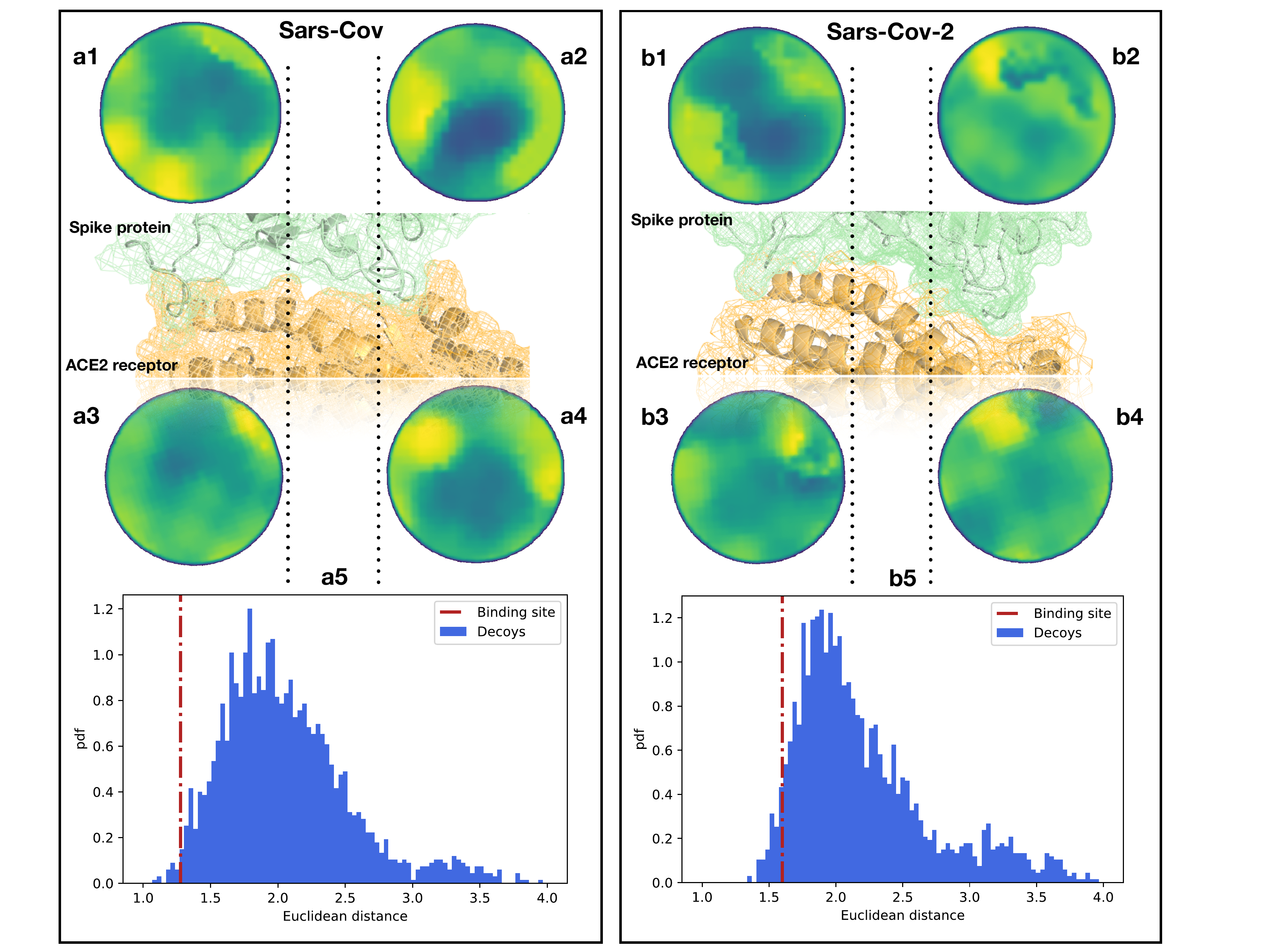}
\caption{\textbf{ Comparison between the binding regions of SARS-CoV and SARS-CoV-2 spike protein with human ACE2.} \textbf{a1,2)} Patch projections in the unitary circle for the two binding regions of Sars-Cov spike protein. \textbf{a3,4)} Patch projections in the unitary circle for the binding regions of the human ACE2 receptor. \textbf{a5)} Distance distribution between ACE2 binding sites and randomly selected patches on the S spike protein of SARS-Cov. Decoys patches are sampled taking two random regions separated by the same distance measured in the experimental structure. The red dotted line represents the distance between the real ACE2 and S spike's patches. \textbf{b)} The same as a) but for the binding site of SARS-Cov-2 and the human ACE2 receptor.  The real distances are in the  1st and 5th  percentiles of the distributions for SARS-CoV and SARS-CoV-2 respectively. 
} 
\label{fig:sars_vs_covid}
\end{figure*}

The excellent and very promising results found in the large dataset encouraged us to investigate the pressing problem of the coronavirus interaction with the host cell.
We first analyzed the shape complementarity between the spike proteins of SARS-CoV and SARS-CoV-2 in complex with human ACE2 receptor \cite{li2005structure,yan2020structural}.
It is interesting to notice that the contact between spike and ACE2 receptor both for SARS-CoV and SARS-CoV-2 occurs in two separate interacting regions (see  Fig.~\ref{fig:sars_vs_covid}), meaning that we need to investigate the two interacting regions separately.
When comparing the two raw distances, we found that ACE2-SARS-CoV  distance is smaller than the ACE2-SARS-CoV-2 one, even if for both complexes the complementarity is much higher than the one would find in other random regions of the complexes (see Figure~\ref{fig:sars_vs_covid}). 
Note that for an appropriate comparison, we need to define a suitable ensemble of random patches. Indeed, the random regions are sampled from the molecular surface of the spike protein imposing that the center of the two patches has the same distance as the binding region observed in the experimental complex. Then, both the real spike binding region and the ensemble of 1000 sampled regions are compared with the receptor binding sites.

The results of this analysis are shown in Fig.~\ref{fig:sars_vs_covid}. We show the distance distribution of the random regions and we report the distance between the real binding regions, both for the Ace2-SARS-CoV and Ace2-SARS-CoV-2 complex.
As the method works in recognizing interacting patches, real binding regions show a higher complementarity (lower distance) than the randomly sampled regions.

Furthermore, this analysis shows that the ACE2 receptor has a slightly higher shape complementarity with SARS-CoV than with SARS-CoV-2 spike protein. However, the results are quite comparable among them, in accordance with experimental data \cite{walls2020structure}.

\subsection*{Identification of another possible binding region of the SARS-Cov-2 spike}
Although it is currently known that the spike protein of SARS-CoV-2 binds to the ACE2 receptor of host cells \cite{wan2020receptor,hoffmann2020sars}, the investigation of possible other infection mechanism is a key factor in the study of this disease. Specifically, in this work \cite{zhou2020pneumonia} the authors underline the necessity to elucidate whether SARS-Cov-2 spike protein could have acquired the ability to bind with sialic acid as  MERS-CoV does.
It has been recently shown that besides the usual receptor (dipeptidyl-peptidase 4 receptor), MERS-CoV spike protein interacts with sialic acid molecules~\cite{park2019structures} using a well-identified pocket in the N-terminal region of the protein. This makes the virus able to interact with high airways and then reach the low airway cells~\cite{Li2017}.

The recognition between spike and sialic acids occurs via a conserved groove which plays a key role for S MERS-CoV-mediated attachment to sialosides and entry into human airway epithelial cells \cite{park2019structures}.

Since the interaction of MERS-CoV spike and the sialic acids is caused mainly by hydrogen bonds and shape complementarity ~\cite{Tortorici2019}, our method is particularly suitable to find 
on the SARS-CoV-2 spike surface a region similar to the one involved in binding of the sialic acid in MERS-CoV spike.

Using the experimental structure of MERS-CoV spike and sialic acid molecules complexes \cite{park2019structures}, we extract its binding region and we describe it through Zernike descriptors. Then we sample the corresponding domains of both SARS-CoV and SARS-CoV-2 spike, building a molecular patch on each surface point and characterizing it with its corresponding Zernike descriptors. Each region sampled from the spike proteins of these 2 viruses is then compared with the MERS-CoV spike binding region, looking for a similar region that can mediate interaction with a similar ligand. 

\begin{figure*}[]
\centering
\includegraphics[width = 0.8\textwidth]{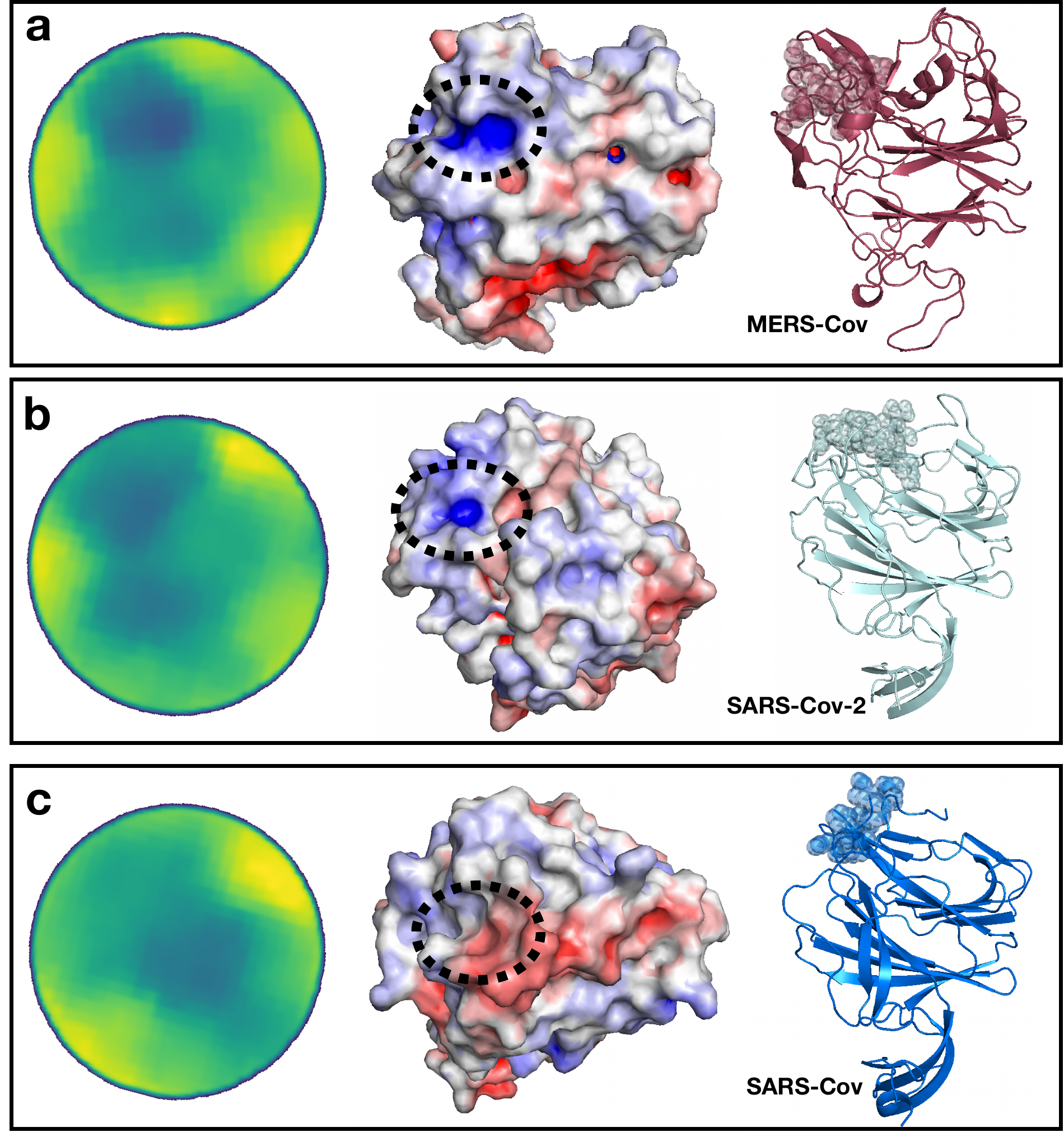}
\caption{\textbf{Identification of a SARS-CoV-2 spike region  very similar to the sialic acid-binding site on MERS-CoV spike.}
\textbf{ a)} From left to right, projected region of the real interaction between MERS-CoV and sialic acid, electrostatic potential surface in the same region and cartoon representation of the MERS-CoV spike protein with highlighted the binding site. 
\textbf{ b)}  Putative sialic acid-binding region on SARS-CoV-2 as predicted by our Zernike-based method. From left to right, the projected region of putative interaction between SARS-CoV and sialic acid, electrostatic potential surface, and cartoon representation of the SARS-CoV spike protein with highlighted the binding site. 
 \textbf{c)} Same as b) but for SARS-CoV spike protein.}
\label{fig:pocket}
\end{figure*}

In Fig.~\ref{fig:pocket}, we show the results of this analysis. In particular, selecting the region most similar to the MERS-CoV binding site we identified one region for both SARS-CoV and SARS-CoV-2 spike protein.

Interestingly the best region of SARS-Cov-2 spike exhibits a higher similarity than the pocket selected by the SARS-CoV spike.
We moreover calculate the electrostatic potential of the involved surfaces with eF-surf web-server \cite{kinoshita2004ef}.
As shown in Fig.~\ref{fig:pocket}, in cartoon representation, the region found in the molecular surface of the SARS-CoV-2 spike is very similar to the MERS spike region interacting with sialic acid, both in terms of shape and electrostatic similarity. Differently, the region identified in the SARS-CoV spike exhibit an electrostatic clearly dissimilar from the sialic binding site in the MERS-CoV spike, making very unlikely the interaction with sialic acid in that region.  

\begin{figure*}[]
\centering
\includegraphics[width = 0.8\textwidth]{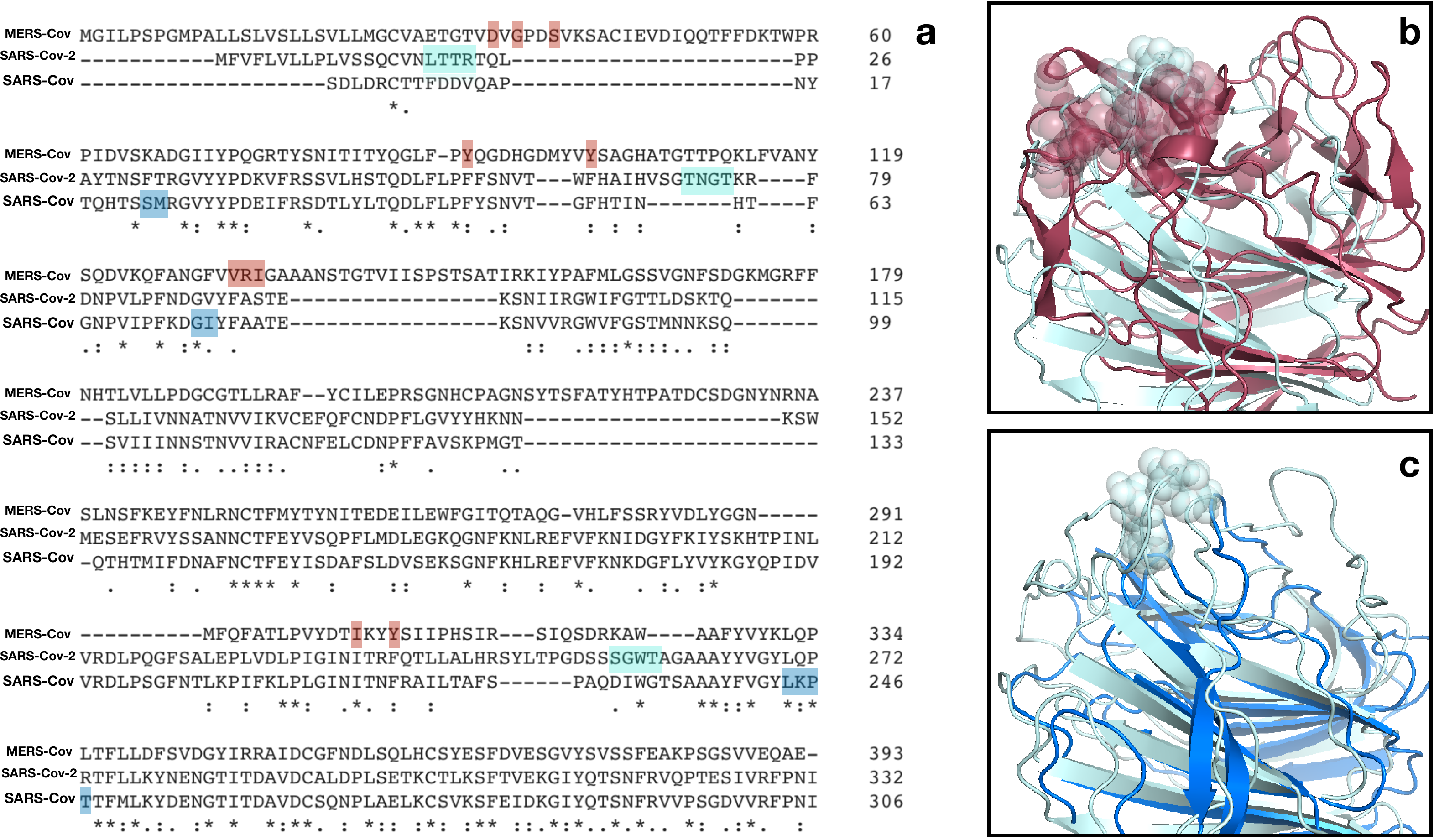}
\caption{\textbf{Sequence and structure comparison of the N-terminal region of MERS-CoV, SARS-CoV-2 and SARS-CoV.}
\textbf{ a)} A multiple sequence alignment between the  MERS-CoV, the SARS-CoV and the SARS-CoV-2 spike protein sequence.\textbf{ b)} Structural comparison between MERS-CoV and SARS-Cov-2 A-domain. The sialic acid binsing site for MERS-Cov spike and the proposed binding site on SARS-Cov-2 spike are highlighted.\textbf{ c)} Structural comparison between SARS-CoV and SARS-Cov-2 A-domain. The proposed binding site on SARS-Cov-2 has no correspondence in SARS-CoV structure.}
\label{fig:sequence}
\end{figure*}

In addition, in Fig.\ref{fig:sequence}, we propose a multiple sequence alignment - with software Clustal Omega\cite{sievers2011fast} - between the three spike proteins. 
 
Importantly the proposed SARS-CoV-2 binding site, besides being structurally in a surface region bordering the corresponding MERS pocket, is composed of a set of consecutive residues (sequence number 73-76) constituting an insertion in respect to SARS-CoV spike sequence. Thus, this insertion in the A-domain of the spike protein could confer to SARS-CoV-2 the capability of infecting human cells in a dual strategy, making possible the high diffusion speed of this virus.  

\section{Discussion}

A blind prediction of the interaction regions between molecules is still an open challenge, despite the great steps that have been made. However, the need for fast and reliable theoretical and computational tools, capable to guide and speed-up experiments, becomes especially important when we face emergencies like the present one. Emergency caused by the novel SARS-CoV-2 human infection is spreading with an impressive rate, such that the World Health Organization officially declared it a pandemic. 

Indeed, the last few months have seen extensive studies about the virus-host interactions focusing in particular on the various stages of the cell internalization mechanism.
Recent works found that, in analogy with the case of SARS-CoV, SARS-CoV-2 uses too its spike protein to bind to ACE2 receptors, mostly associated in the lower respiratory ways. 
Further experimental investigations highlighted a comparable receptor binding affinity between the novel coronavirus and the older SARS-CoV, even if the binding regions display a certain degree of variability~\cite{andersen2020proximal}. The modest difference in binding affinity seems insufficient to explain the higher human-human transmission rate with respect to SARS-CoV and the overall sequence variability suggest that SARS-CoV-2 may have optimized in other directions, such as in acquiring the ability of binding to other receptors~\cite{zhou2020pneumonia}. 

In this work, we present a new fast computational method that compactly summarizes the morphological properties of a surface region of a protein. Testing the unsupervised method on a large dataset of protein-protein interactions, we proved its excellent ability to correctly recognize the high shape complementarity occurring between interacting surfaces. 
Analyzing the newly available experimental structures of SARS-CoV-2 Spike protein in complex with human ACE2, we found that the binding region presents indeed a comparable  (slightly slower)  shape complementarity with the analogous complex of SARS-CoV.
Such a minimal difference enforces the hypothesis that the apparent higher fitness of SARS-CoV-2 should lie elsewhere.

In particular, looking at other members of the large coronavirus family, one finds that many members developed the ability to bind to two distinct receptors, with one binding site in the C-terminal domain of the S-protein that generally binds protein-like receptors (like ACE2 for SARS-CoV and SARS-CoV-2)  and the other situated in the N-terminal region, usually able to bind to sugar-like receptors.
In particular, MERS-CoV has been found able to bind to sialic acid receptors both in camel, human and bat cells.
Applying our method to the sialic acid-binding region, which has been recently determined experimentally in MERS-CoV, we have found an exceptionally similar region in the corresponding region of the SARS-CoV-2 spike. This region, similar in structure to the MERS-CoV correspondent one and absent in SARS-CoV (see Figure~\ref{fig:sequence}), could be able to mediate a low-affinity but high-avidity interaction with sialic acid. 
Interestingly, the sequence variability of the spike protein, recently studied considering  SARS-CoV-2 sequences belonging to 62 different strains ~\cite{2003.13655}, shows a high conservation level of the ACE2 binding site while the highest variability is located in the region that we indicate here to be potentially involved in sialic acid biding: this evidence confirm the importance of this region in regulating host-cell infection \cite{qing2020distinct}.

In conclusion, we propose that this dual cell entry mechanism can explain the high diffusion speed this virus exhibits and we strongly encourage a more accurate investigation of this observation.

\section{Methods}

\subsection{Datasets}

\subsubsection*{Protein-Protein Dataset}

A dataset of protein-protein complexes experimentally solved in x-ray crystallography is taken  from~\cite{gainza2020deciphering}.

We only selected pair interactions regarding chains with more than 50 residues. The Protein-Protein dataset is therefore composed of 4634 complexes.

\subsubsection*{Experimental protein structures}
\begin{itemize}
    \item Complex between SARS-CoV spike protein and human ACE receptor: PDB code 6ACJ.
    \item Complex between SARS-CoV-2 spike protein and human ACE receptor: PDB code 6M17.
    \item Complex between MERS spike protein and sialic acid: PDB code 6Q07.
    \item Unbound SARS-CoV spike protein: PDB code 6CRV.
    \item Unbound SARS-CoV-2 spike protein: modeled by I-TASSER server~\cite{yang2015tasser}. 
\end{itemize}

\subsection{Computation of molecular surfaces}   
We use DMS~\cite{richards1977areas} to compute the solvent accessible surface for all proteins structure, given their x-ray structure in PDB format~\cite{berman2003protein}, using a density of 5 points per $\AA^2$ and a water probe radius of 1.4 $\AA$. The unit normals vector, for each point of the surface, was calculated using the flag $-n$.

\subsection{Patch selection and space reduction}

Given a molecular surface described as a set of point in a three-dimensional Cartesian space, and a region of interest on this surface,
we define a surface patch, $\Sigma$ as the intersection of the region of interest and the surface.
In principle, the region of interest can have an arbitrary shape, in this work we chose to use a spherical region having radius $R_s = 6 \AA$ and one point of the surface as the center (see Figure~\ref{fig:computational_protocol}a).
Once the patch is selected, we fit a plane that
passes through the points in $\Sigma$, and we reorient the patch in such a way to have the z-axis perpendicular to the plane and going through the center of the plane. Then given a point $C$ on the z-axis
we define the angle $\theta$ as the largest angle between the perpendicular axis and a secant connecting $C$ to any point of the surface $\Sigma$.
$C$ is then set in order that $\theta=45^\circ$.
$r$ is the distance between $C$ and a surface point. We then build a square grid, and we associate each pixel with the mean $r$ of the points inside it. This 2D function can be expanded in the basis of the Zernike polynomials: the norm of the coefficients of this expansion constitute the Zernike invariant descriptors,  which are invariant under rotation in the space. 
In the next section, we provide a brief summary of the main features of the Zernike basis. There are several good reviews,  like~\cite{lakshminarayanan2011zernike} that offer more detailed discussions. A schematic representation of the procedure is shown in Figure~\ref{fig:computational_protocol}b-d.

\subsection{2D Zernike polynomials and invariants}

Given a function $f(r,\phi)$ (polar coordinates) defined inside the region $r < 1$ (unitary circle), it is possible to represent the function in the Zernike basis as 
\beq
f(r,\phi) = \sum_{n=0}^{\infty} \sum_{m=0}^{m=n} c_{nm} Z_{nm}
\eeq

with 

\begin{multline}
c_{nm} =  \frac{(n+1)}{\pi} \braket{Z_{nm}|f} =\\= \frac{(n+1)}{\pi} \int_0^1 dr r\int_0^{2\pi} d\phi Z_{nm}^*(r,\phi) f(r,\phi).
\end{multline}

being the expansion coefficients.
The Zernike polynomials are complex functions, composed by a radial and an angular part,

\beq
Z_{nm} = R_{nm}(r) e^{im\phi}.
\eeq

where the radial part for a certain couple of indexes, $n$ and $m$, is given by

\beq
R_{nm}(r) = 
\sum_{k= 0}^{\frac{n-m}{2}} \frac{(-1)^k (n-k)!}{k!\left(\frac{n+k}{2} - k\right)!\left(\frac{n-k}{2}-k\right)!} r^{n-2k}
\eeq

In general, for each couple of polynomials, one finds that

\beq
\braket{Z_{nm}|Z_{n'm'}} = \frac{\pi }{(n+1)}\delta_{nn'}\delta_{mm'}
\eeq

which ensures that the polynomials can form a basis and 
knowing the set of complex coefficients, $\{ c_{nm} \}$ allows for a univocal reconstruction of the original image (with a resolution that depends on the order of expansion, $N = max(n)$). We found that $N=20$, which corresponds to a number of coefficients of 121, allows for a good visual reconstruction of the original image.

By taking the modulus of each coefficient ($z_{nm} = |c_{nm}|$), a set of descriptors can be obtained which have the remarkable feature of being invariant for rotations around the origin of the unitary circle.

The shape similarity between two patches can then be assessed by comparing the Zernike invariants of their associated 2D projections.
In particular, the similarity between patch $i$ and $j$ is measured as
the Euclidean distance between the invariant vectors, i.e.
\beq 
d_{ij} = \sqrt{\sum_{k=1}^{M=121} (z_i^k - z_j^k)^2}
\eeq

\subsection{Evaluation of similarity and complementarity}
When comparing patches, the relative orientation of the patches before the projection in the unitary circle must be evaluated. Intuitively, if we search for similar regions we must compare patches that have the same orientation once projected in the 2D plane, i.e. the solvent-exposed part of the surface must be oriented in the same direction for both patches, for example as the positive z-axis. If instead, we want to assess the complementarity between two patches, we must orient the patches contrariwise, i.e. one patch with the solvent-exposed part toward the positive z-axis (`up') and the other toward the negative z-axis (`down').

\subsection{Blind search of binding sites}
The velocity of the procedure that from a patch in the 3D surface produces the set of invariant descriptors, allows for a vast screening of couples of surfaces to look for both similar and also complementary regions. 
In order to identify the binding region given a couple of proteins, we can associate to each point of one surface a vector of  Zernike invariants associate to the `up' patch having that point as center and another set of invariants to each point of the other surface ( with `down' orientation).
Then for each point $i$ of say, protein 1, we can compute the Euclidean distance with all the points of the other surface associated with protein 2 and associate to point $i$  the minimum found distance and vice-versa for protein 2 (see Figure~\ref{fig:computational_protocol}e-f).
A smoothing process of the surface point values is applied in order to highlight the signal in the regions characterized mostly by low distance values (high shape complementarity).

\newpage

\section*{Acknowledgment}

The authors would like to thank Prof. Gian Gaetano Tartaglia for very helpful discussions.


\end{document}